\documentclass[cits]{JINST}
\usepackage{amsmath}
\title{Dependence of the energy resolution of a scintillating crystal 
on the readout integration time.}

\author{V. Bocci$^a$, D. Chao$^c$, G. Chiodi$^a$, R. Faccini$^{a,b}$,  
F. Ferroni$^{a,b}$, R. Lunadei$^a$,  G. Martellotti$^{a}$, G. Penso$^{a,b}$, 
D. Pinci$^{a}$, L. Recchia$^{a}$\\
\llap{$^a$} INFN Sezione di Roma, Roma, Italy \\
\llap{$^b$} Dipartimento di Fisica, Sapienza Universit\`a di Roma, 
Roma, Italy \\
\llap{$^c$} California Institute of Technology, Pasadena, California, USA\\
E-mail: \email{riccardo.faccini@roma1.infn.it}}

\abstract{
The possibilty of performing high-rate calorimetry with a slow scintillator 
crystal is studied. In this experimental situation, to avoid pulse pile-up, 
it can be necessary to base the energy measurement on only a fraction of 
the emitted light, thus spoiling the energy resolution.
This effect was experimentally studied with a 
BGO crystal and a photomultiplier followed by an integrator,
by measuring the maximum amplitude of the signals. 
The experimental data show that the energy resolution is
exclusively due to the statistical fluctuations of the
number of photoelectrons contributing to the maximum amplitude.
When such number is small its fluctuations are even smaller
than those predicted by Poisson statistics. 
These results were confirmed by a Monte 
Carlo simulation which allows to estimate, in a general 
case, the energy resolution, given the total number of 
photoelectrons, the scintillation time and the integration time.}

\begin{document}
\vspace*{3truemm}

\section{Introduction}
Electromagnetic calorimeters are often composed of inorganic scintillating
crystals viewed by photodetectors.
The energy resolution attainable depends primarily 
on the number of optical photons emitted by a scintillating crystal for 
a given energy deposit. Usually 
only a fraction of these photons are collected by a photodetector. 
In the following we consider a photomultiplier (PM), 
where the collected optical photons are converted in photoelectrons
with an efficiency characteristic of the photocathode and amplified up to
the anode by the dynode system.
 
Besides a possible non-linear crystal response~\cite{dorenbos}, the anode
charge pulse is therefore proportional to the energy deposited in the crystal 
by the primary particle and the fluctuations of this charge determine
the energy resolution of the detector. 
The main contribution to this resolution comes from the statistics of the
photoelectrons. In addition a small but not negligible contribution comes 
from the fluctuations in the PM gain and in particular from the gain of
the first dynode. This resolution is worsened if the readout 
electronics integrates only part of the total charge delivered
by the anode of the PM. This eventuality can occur when scintillating 
crystals with a 
long decay time are used in high-rate experiments where short 
integration times are needed.

In the present paper we will consider the possibility of using
BGO crystals and PM's for high rate experiments. 
In that case to limit the pile-up and the dead time, the PM 
pulses must be integrated over a time interval shorter than the scintillation 
time of the BGO ($\tau_{scint} = 300$~ns). The deposited energy is 
then deduced from the maximum amplitude of the integrated 
pulse, measured by a peak-sensitive circuit. 
With this procedure only a fraction of the total charge is measured and a
faster response is obtained at the cost of a worse resolution.
This effect was 
experimentally studied for different integration times and 
the results were compared with a Monte Carlo (MC) simulation.

\section{Experimental setup}
\label{sec:exp}
\begin{figure}[hb]
\begin{center}
\includegraphics[width =.75 \textwidth] {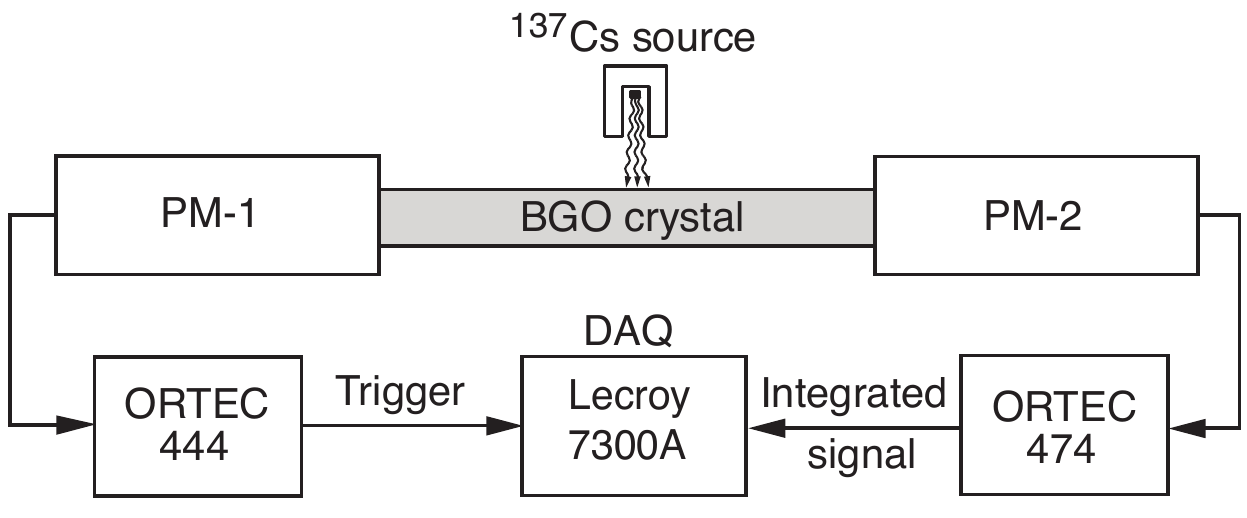}
\caption{\small{Experimental setup.}}
\label{figsetup}
\end{center}
\end{figure}

The experimental setup is sketched in Fig.~\ref{figsetup}.
Photons with an energy of 662~keV from a $^{137}$Cs radioactive source 
are detected by a $2 \times 2\times 18$ cm$^3$ BGO crystal, read 
at both ends by two EMI-9814B PMs.
\vskip 3truemm 
One of the two photomutipliers (PM-1) was used to trigger the 
acquisition of 
the pulses from the other photomultiplier (PM-2).
In order to get rid of the noise,
the trigger-signal from PM-1 was amplified and shaped with a gated biased amplifier (ORTEC-444)
having an integration time of 250 ns.
The signal from PM-2 was processed by a filter amplifier (ORTEC-474) that
has a variable gain and an integration\footnote{The differentiation 
control of the 474 module was set in the $out$ position. This corresponds 
to a differentiation time of 0.2~ms which has a negligible effect on the 
output signals.} time ($\tau_{int}$) that can be
set to 20, 50, 100, 200 and 500~ns. The output signal of that module was 
then acquired 
by a Lecroy WavePro 7300A digital oscilloscope, having a bandwidth 
of 300 MHz and a sampling rate of 250 MS/s.

\section{Equivalent circuit}
\label{equicircuit}

\begin{figure}[t]
\begin{center}
\includegraphics[width =.7 \textwidth] {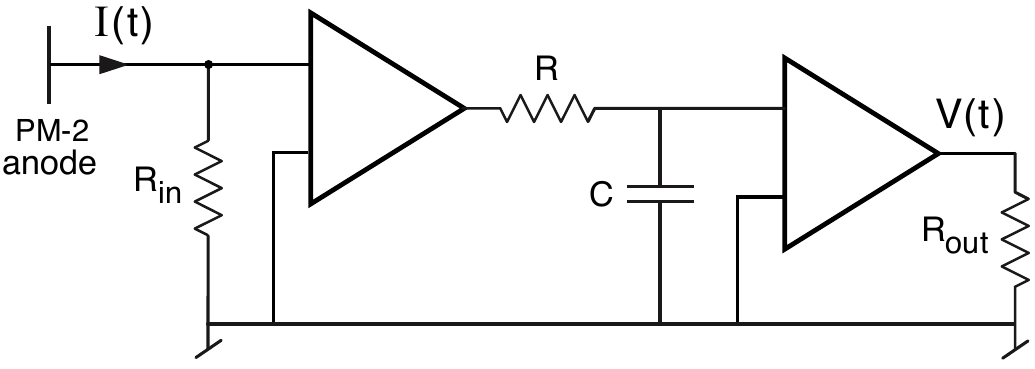}
\caption{\small{Equivalent circuit of the ORTEC-474 integrating 
amplifier. The two buffer amplifiers have an overall gain G while the
integration time $\tau_{int}$ is equal to RC. }}
\label{amplifier}
\end{center}
\end{figure}

The ORTEC 474 integrating amplifier and its connection to the PM-2 
anode can be represented by the equivalent circuit reported in 
Fig.~\ref{amplifier}.
When a given energy is released in the crystal at 
the time $t=0$, the anode current\footnote{The formulas reported
in this section 
hold for a very large  number of photoelectrons per pulse, so 
that the statistical fluctuations are negligible.} delivered by PM-2 is:
\vskip 2truemm
\begin{equation}
I(t) = I_0 \; e^{-t/\tau_{scint}} \times u(t)
\label{eqexp}
\end{equation}
\vskip 3truemm
\noindent
where $\tau_{scint}$ is the decay time of the scintillator
and $u(t)$ is the unit step function.
The total charge per pulse 
released by the anode of PM-2 is \hbox{$Q = I_0 \; \tau_{scint}$}. 
Using the standard Laplace transform method,
the output signal $V(t)$ of the circuit on Fig.~\ref{amplifier}
turns out to be:
\vskip 3truemm
\begin{equation}
V(t) = \frac{G \; I_0 \; R_{in}}{1-\alpha} 
\; (e^{-t/\tau_{scint}}-e^{-t/\tau_{int}})
\label{eqvt}
\end{equation}
\vskip 3truemm
\noindent where $R_{in}$ is the input resistance of the first buffer 
amplifier, $G$ is the overall gain, 
$\tau_{int} = RC$ is the 
integration time and $\alpha=\tau_{int} /\tau_{scint}$.
For $N_{pe} \to \infty$, the maximum amplitude of the output signal is:
\vskip 3truemm
\begin{equation}
A_{N_{pe} \to \infty} = I_0 \; R_{in} \; 
\alpha^{\;\; \alpha/(1- \; \alpha)}
\label{eqA}
\end{equation}
\vskip 3 truemm
\noindent
which occurs at the time:
\vskip -1 truemm
\begin{equation}
T_{N_{pe} \to \infty} =\tau_{int} \; 
\frac{ ln \, \alpha}{\alpha-1}
\label{eqT}
\end{equation}
\vskip 4 truemm
\noindent
From Eq.~\ref{eqvt} the total charge ($Q_{out}$) of an output pulse 
is proportional to $Q$:
\vskip 3truemm
\begin{equation} 
Q_{out}= \int_0^\infty \! \frac{V(t)}{R_{out}} \, dt 
= G \; I_0 \; R_{in}\;\tau_{scint}= G\;R_{in}\;Q
\label{eqQout}
\end{equation}

\section{Data taking}

\begin{figure}[t]
\begin{center}
\includegraphics [width =0.95 \textwidth] {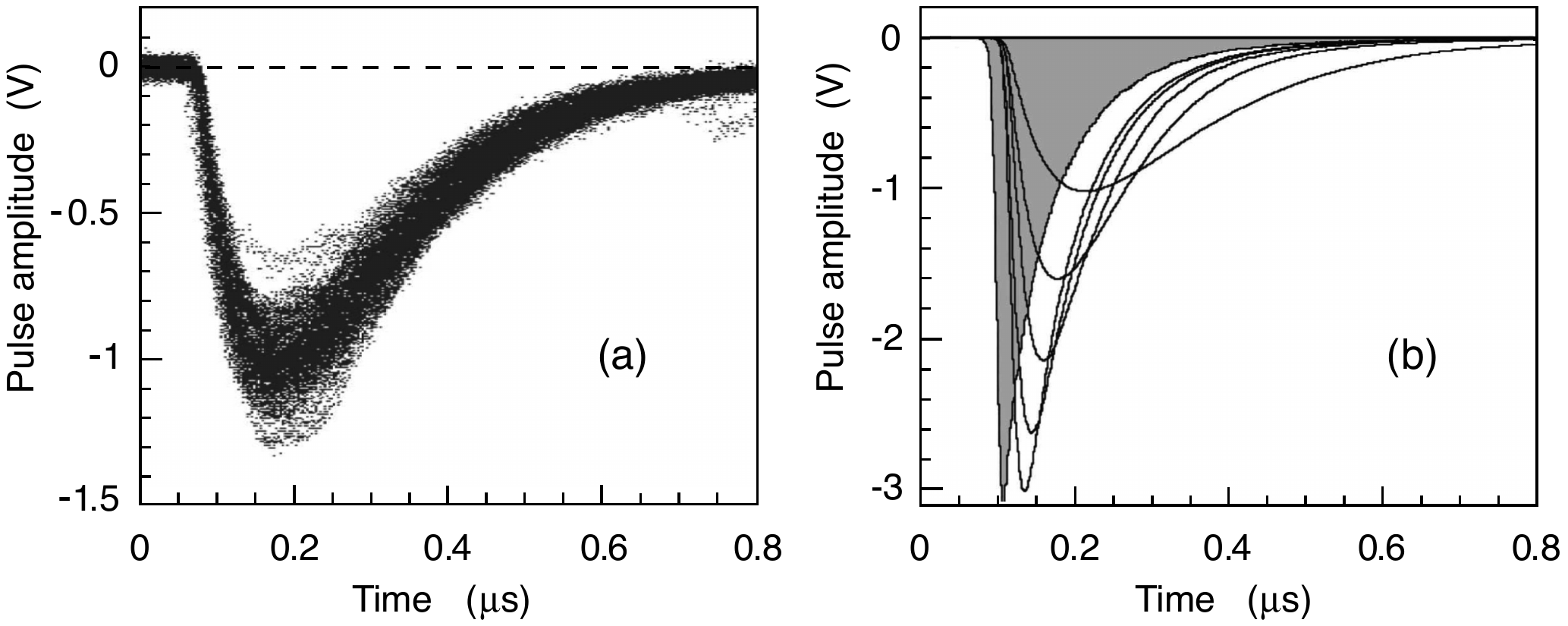}
\caption{\small{
(a): A thousand of superimposed waveforms acquired with $\tau_{int}=$
500~ns. 
(b): Waveforms obtained by averaging 15000 pulses: without integration 
(grey area not normalized)
and with an integration time $\tau_{int}$ = 20, 50, 100, 
200 and 500~ns respectively from the larger to the smaller signal.
}}
\label{PULSES}
\end{center}
\end{figure}

About 15000 pulses were recorded for each of the five possible  values
$\tau_{int}=$ 20, 50, 100, 200 and 500~ns. Each pulse 
was sampled every 4~ns during 10~$\mu$s and the 
resulting 2500 istantaneous amplitudes were acquired. 
A thousand pulses are shown in Fig.~\ref{PULSES}a, 
while in Fig.~\ref{PULSES}b the average waveform is 
reported for the five $\tau_{int}$ values. 

To check the reproducibility of the results, 
four data-sets ({\bf A, B, C} and {\bf D}) were acquired under different 
conditions for all values of $\tau_{int}$.
\begin{itemize}
\item 
{\bf A} and {\bf B} were taken under
the same conditions (HV of PM-2 equal to 1800~V and gain of the 
ORTEC-474 set to 10) but
in different days in order to test the reproducibility of the results;
\item 
{\bf C}~: HV of PM-2 equal to 1800~V and the ORTEC-474 gain 
set to 2;
\item 
{\bf D}~: HV of PM-2 equal to 1750~V  and 
the ORTEC-474 gain set to 10. 
\end{itemize}
During the acquisition of the four data-sets the configuration on the 
trigger side of the set-up (\hbox{PM-1} and ORTEC 444) was kept fixed.
The trigger level of the oscilloscope was set sufficiently low to accept 
all the pulses from the 662 keV photons. 
The electronics noise was measured by acquiring data with a random 
triggers. The width of these noise spectra being 10 times smaller 
than that of the source signal, the contribution of the 
electronics noise to the energy resolution was neglected.

\section{Data analysis}
\vskip 2truemm
The acquired waveforms were analysed off-line. 
For each value of $\tau_{int}$ and for each pulse,
the maximum amplitude ($A$), the peaking time ($T$) and the total charge 
($Q_{out}$) were evaluated. \vspace*{3truemm}

\subsection{Total charge fluctuations}
\label{sectotch}
\begin{figure}[t]
\begin{center}
\includegraphics[width =.9\textwidth] {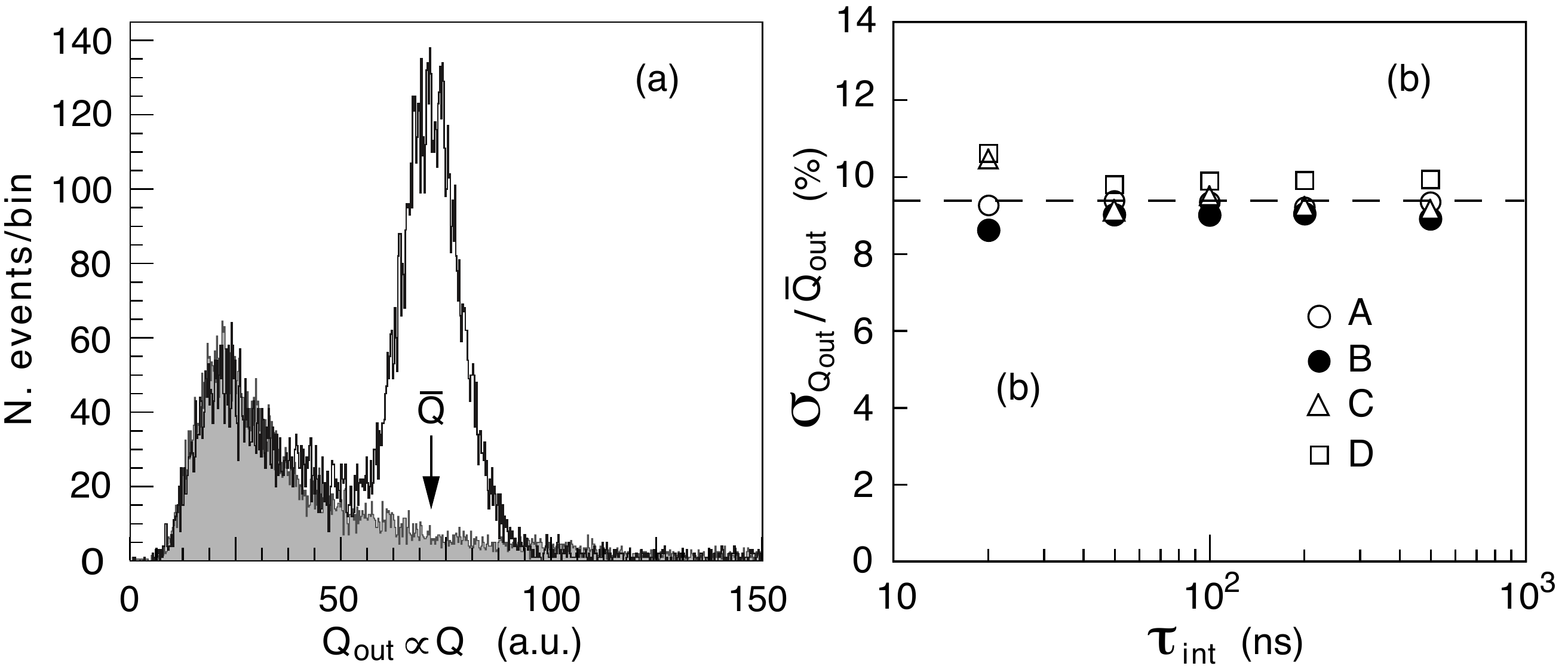}
\caption{\small{(a): A typical spectrum of the total pulse charge.
The peak from 662~keV gammas is superimposed to a background (grey area) 
which was measured without the $^{137}$Cs source.
(b): Relative total charge resolution as a function of 
$\tau_{int}$ for the four sets (A, B, C and D) of data taking. The 
dashed line represents the expected behaviour for 
$\overline{N}_{pe} = 130$. The statistical errors are within 
the points. 
}}
\label{figsigmaQsuQ}
\end{center}
\vskip -6truemm
\end{figure}

\vskip 2truemm
When the total charge of each pulse is measured, the energy 
resolution of the BGO crystal is 
mainly determined by the statistical 
fluctuations of the total number ($N_{pe}$) of the photoelectrons 
and by the fluctuations of the gain of the 12 PM dynodes ($g_1,...,g_{12}$).
Assuming a Poisson distribution for $N_{pe}$ and considering that 
in general $g_1\;>g_2\;=\;g_3\;=\;...\;=\;g_{12}\;\equiv\;g$
the energy resolution\footnote{Throughout this paper $\overline{x}$ 
and $\sigma_x$  indicate respectively the mean value and the r.m.s. of 
a Gaussian fit to the $x$ distribution.} is given by \cite{RTC,PM1,PM2,PM3,PM4}:
\vskip 5truemm
\begin{equation}
\frac{\sigma_E}{\overline{E}} = \frac{\sigma_Q}{\overline{Q}} = 
\frac{\sigma_{Q_{out}}}{\overline{Q}_{out}} =
\frac{1}{\sqrt{\overline{N}_{pe}}} \; 
\sqrt{1+\frac{1}{\overline{g}_1}
\left(\sum\limits_{i=0}^{k-1}\frac{1}{\overline{g}^{\;i}} \right)} \simeq
\frac{1}{\sqrt{\overline{N}_{pe}}} \; 
\sqrt{1+\frac{1}{\overline{g}_1}
\left(\frac{1}{1-1/g}\right)}
\label{eqsigmaEsuEOrig}
\end{equation}
\vskip 6truemm
\noindent
where $E$ is the measured energy and
$k\;=\;12$ is the number of dynodes.
Since $g$ is rather larger than 1, 
Eq.\ref{eqsigmaEsuEOrig} was approximated as:
\begin{equation}
\frac{\sigma_E}{\overline{E}} = \frac{\sigma_Q}{\overline{Q}} = 
\frac{\sigma_{Q_{out}}}{\overline{Q}_{out}} \simeq
\frac{1}{\sqrt{\overline{N}_{pe}}} \; 
\sqrt{1+\frac{1}{\overline{g}_1}}
\label{eqsigmaEsuE}
\end{equation}
that is equivalent to take into account only the contribution of the 
fluctuations of the first dynode.
In Fig.~\ref{figsigmaQsuQ}a a typical experimental spectrum of $Q_{out}$ 
is shown. 
Taking into account the HV of PM-2, the 
characteristics of the 9814B tube \cite{ET} and of the BeCu dynodes 
\cite{RTC} the average gain of the first dynode was assumed to 
be $\overline{g}_1\simeq 6$. \\
\noindent The energy resolution becomes:
\vskip 1 truemm
\begin{equation}  
\frac{\sigma_E}{\overline{E}} =
\frac{\sigma_{Q_{out}}}{\overline{Q}_{out}} =
\frac{1.08}{\sqrt{\overline{N}_{pe}}}
\label{eqNpe}
\end{equation}
\vskip 1 truemm
In Fig.~\ref{figsigmaQsuQ}b the
experimantal values of $\sigma_{Q_{out}}/\overline{Q}_{out}$ are reported 
for the five
values of $\tau_{int}$ and for the four sets of data taking.
They show, as expected, a flat
behaviour with respect to $\tau_{int}$. A constant fit to these data 
allows to determine from Eq.~\ref{eqNpe} the average number of 
photoelectrons $\overline{N}_{pe} = 130^{+13}_{-11}$.
\vskip 5 truemm

\subsection{Maximum amplitude fluctuations}
\label{secmaxamp}
\begin{figure}[t]
\begin{center}
\includegraphics[width =.9\textwidth] {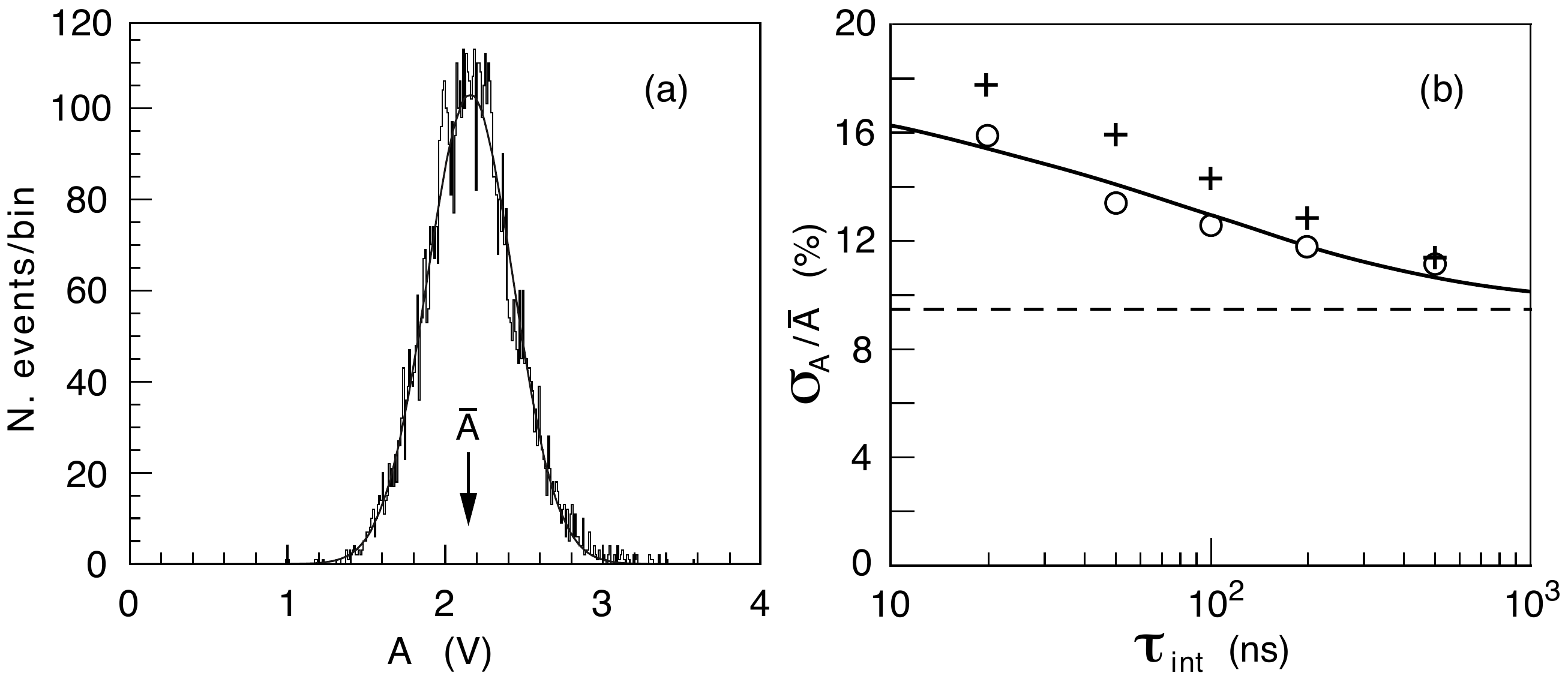}
\caption{\small{
(a): Example of a spectrum of the maximum pulse amplitude A. 
The curve is a gaussian fit to the data.
(b): Experimental maximum amplitude resolution (open points) as a function 
of $\tau_{int}$. The crosses are the 
values predicted by 
Eq.'s~\protect\ref{eqnpe} and \protect\ref{approxnpe} 
with $\overline{N}_{pe}=130$ and with the measured values of 
$\overline{T}$ (Fig.~\protect\ref{T}). The statistical errors are 
within the points. 
The curve is the prediction of the MC (see Sec.\protect\ref{secMC}).
The dashed line is the asymptotic value of the curve, 
which is equal to $\sigma_{Q_{out}}/\overline{Q}_{out}$.}} 
\label{figsigmaAsuA}
\end{center}
\end{figure}

\vskip 3truemm
When the measurement of the total charge $Q$ takes 
too long, the energy deposited in the crystal can be inferred
from the maximum amplitude $A$ of the integrated signal. 
In Fig.~\ref{figsigmaAsuA}a
a spectrum of $A$ obtained in the present test with 
$\tau_{int} = 100$~ns is shown.
The fluctuations ($\sigma_A$) on $A$ are obtained
by a gaussian fit to the data.
In Fig.~\ref{figsigmaAsuA}b the experimental resolution
$\sigma_A/\overline{A}$ is reported\footnote{The data from 
the four data sets A, B, C and D have been averaged.}
as a function of $\tau_{int}$.
At large values of $\tau_{int}$
and in particular for $\tau_{int} \gg \tau_{scint}$, 
$\sigma_A/\overline{A}$ 
tends to the value of $\sigma_{Q_{out}}/\overline{Q}_{out}$, while
at smaller values of $\tau_{int}$ the resolution worsen.
To clarify the dependence of $\sigma A / A$ on $\tau_{int}$, 
a naive Poissonian model based on an extension of 
Eq.~\ref{eqNpe} was 
adopted. According to this model the resolution is given by: 
\begin{equation}
\frac{\sigma_E}{\overline{E}}=
\frac{\sigma_A}{\overline{A}} =
\frac{1.08}{\sqrt{\overline{n}_{pe}}}
\label{eqnpe}
\end{equation}
%
\noindent where $n_{pe}$ is the number of 
photoelectrons which contribute, for each event, to its maximum amplitude 
$A$, i.e. those emitted before the peaking time $T$ of that event.

\begin{figure}[t]
\begin{center}
\includegraphics[width =.5\textwidth] {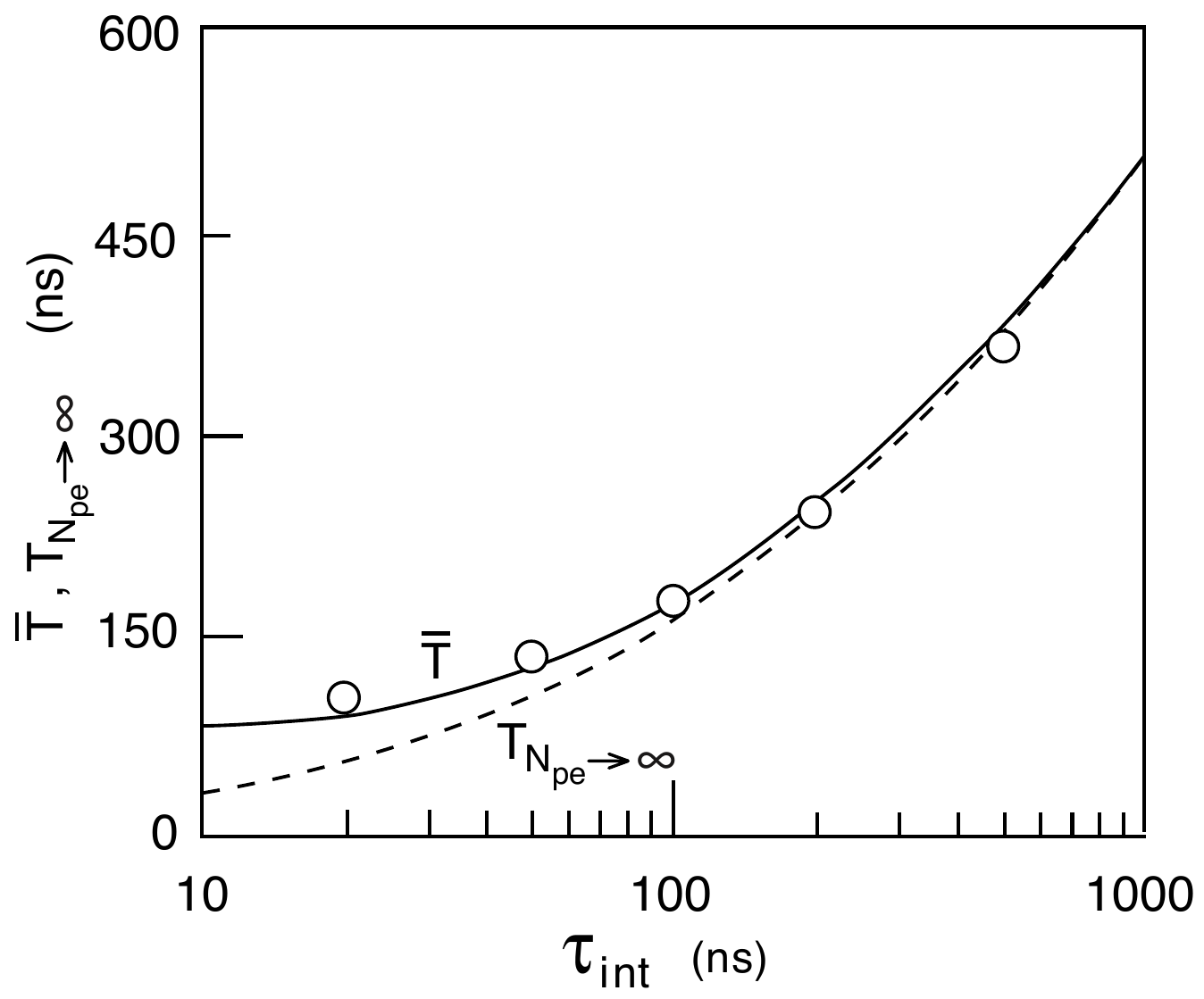}
\caption{\small{Experimental average peaking time 
($\overline{T}$) for the five considerd values of $\tau_{int}$.
The error bars are within the points. 
The continuous curve is the prediction of the MC (see 
Sec.\protect\ref{secMC}) for $\overline{N}_{pe} = 130$. When 
\hbox{$N_{pe} \to \infty$} the continuous curve tends to the 
dashed curve represented by Eq.~\protect\ref{eqT}. 
}}
\label{T}
\end{center}
\end{figure}
The average value $\overline{n}_{pe}$ was
approximated to the fraction $F$ of the average 
total number of photoelectron emitted before the 
experimentally measured values of $\overline{T}$: 
\begin{align}
{\overline{n}_{pe}}& = F \overline{N}_{pe} \\
F& = (1 - e^{-\overline{\vphantom{|^.}T}/\tau_{scint}})             
\label{approxnpe}
\end{align}
 In Fig.~\ref{T} the measured dependence of 
$\overline{\vphantom{|^.}T}$ on $\tau_{int}$ is reported.

In Fig.~\ref{figsigmaAsuA}b the predictions of Eq.'s~\ref{eqnpe} and
\ref{approxnpe} are
compared with the experimental results. While for 
$\tau_{int} = 500$~ns the agreement is good, at lower values of 
$\tau_{int}$ the experimental resolution is better
than predicted with the naive model. To understand this 
discrepancy, a detailed Monte Carlo simulation of the 
experimental situation was performed.
\vspace*{5truemm}

\section{The Monte Carlo simulation}
\label{secMC}

The formulae reported in Sec.~\ref{equicircuit} represent the 
response of an RC integrator excited by an 
exponentially decreasing current composed of
a very large number of electrons, so that the charge quantization
is washed out. In the situation we are considering, the average
number of photoelectrons per pulse is relatively small so that
the response of the device must be simulated with a MC.
The input current $I(t)$ is described as 
a sum of delta functions, each one corresponding to an incoming 
photoelectron. Then the amplitude of the integrated pulse at a time $t$ 
turns out to be the 
sum of the contributions from all the photoelectrons emitted before that 
time\footnote{It can be shown that
\hbox{Eq.~\ref{eqpulseshape} $\to$ Eq.~\ref{eqvt}} when 
\hbox{$N_{pe} \to \infty$.}}: 
\begin{equation}
V(t) = R_{in} \; (\frac{q}{\tau_{int}}) 
\sum\limits_{i=1}^{n(\,t)}G_i \; e^{-(t-t_i)/\tau_{int}}
\label{eqpulseshape}
\end{equation}
\vskip 8pt
\noindent
where $q$ is the electron charge, $n(t)$ is the number of 
photoelectrons 
emitted before the time $t$, $t_i$ ($i=1,n$) is the emission time 
of the $i$-th electron ($0 < t_i < t$) and $G_i$ is the PM 
gain for the $i$-th electron.
The probability distribution function of the $t_i$ is a
decreasing exponential with a decay time equal to $\tau_{scint}$. 
For a fixed $\; t$, $n(t)$ follows a Poisson distribution with a mean
\begin{equation}
\overline{n}(t) = \overline{N}_{pe} ( 1 - e^{-t/\tau_{scint}})
\label{mean}
\end{equation}

It is worthwile to note that Eq.~\ref{eqpulseshape} represents a
single pulse that can be used to measure the energy, only if there is an 
effective pile-up of the contributions
of many photoelectrons belonging to the same detected particle. This 
occurs if the integration time 
$\tau_{int}$ is much larger than the average time interval between two 
consecutive photoelectrons ($\thickapprox \tau_{scint}/\overline{N}_{pe}$): 
\begin{equation}
K \equiv \overline{N}_{pe} \frac{\tau_{int}}{\tau_{scint}} \gg 1
\label{k}
\end{equation}
\noindent while if $K \lesssim 1$ the energy released in the 
scintillator gives rise only to a series of single photoelectron pulses.
In the present experiment $K$ ranges from 8.7 (at $\tau_{int}=20$~ns) to
217 (at $\tau_{int}=500$~ns). 

In the MC simulation all the aforementioned effects were taken into 
account. The MC was run 
with $10 \leq \overline{N}_{pe} \leq 10^4$ and
for $\tau_{int}$ ranging from 10~ns to 1~$\mu$s. 
For each of the $\sim 100$ pairs of values 
$(\overline{N}_{pe},\tau_{int})$ about 
10000 pulses were generated.
For each simulated pulse the MC calculates its amplitude $V(t)$ every 
ns during an interval of 2 $\mu$s. The maximum amplitude $A$
and the time $T$ at which this maximum occurs 
were recorded for each pulse and the relative fluctuations 
$\sigma_A/\overline{A}$ were determined.
\vspace*{5truemm}

\subsection{Comparison with experimental data}
For a comparison with the experimental data, the MC was 
run with $\overline{N}_{pe} = 130 $. 
The Poisson fluctuations of the gain $g$ of the first dynode,
with a mean value $\overline{g} = 6$, were also taken into account.

\begin{figure}[t] 
\begin{center}
\includegraphics[width =.85\textwidth] {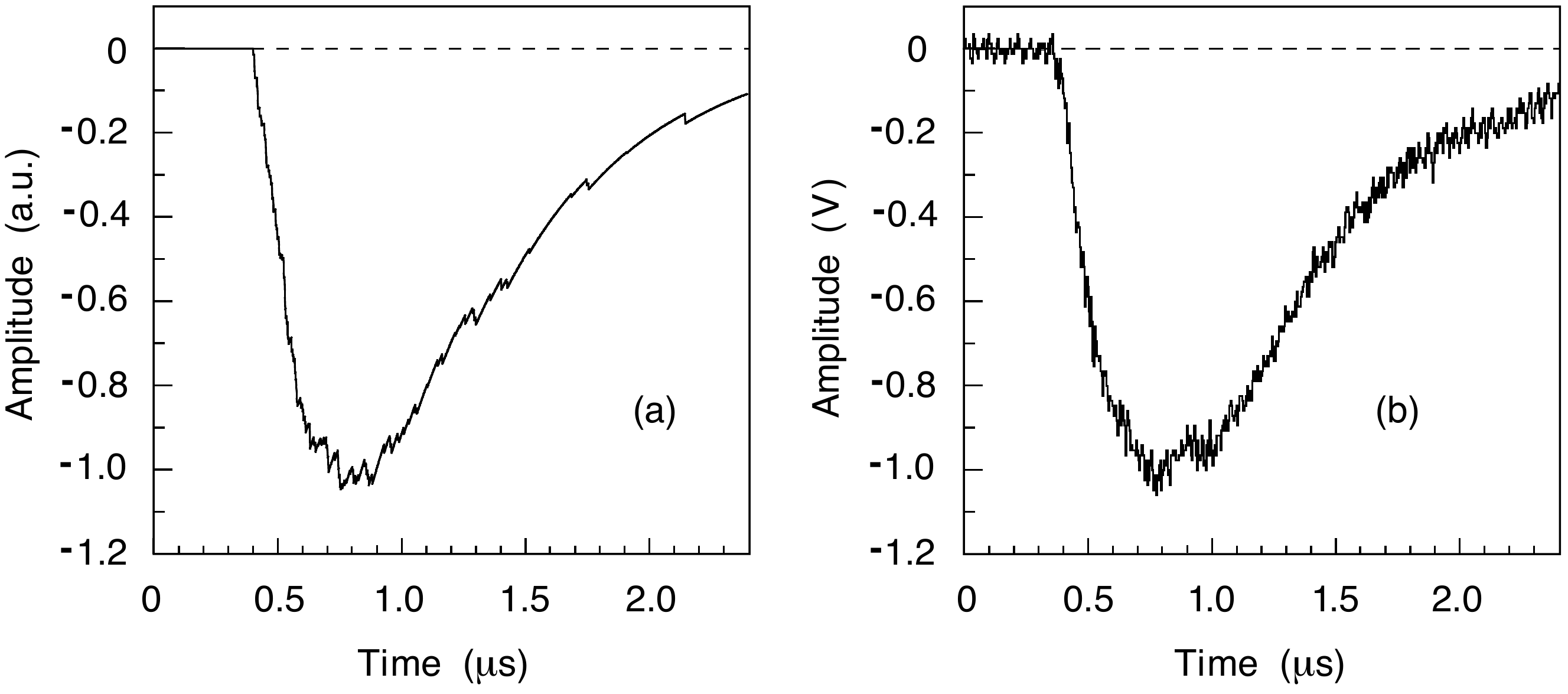}
\caption{\small{(a) A typical pulse generated by MC 
with $N_{pe}=130$ and $\tau_{int}= 500$~ns. 
Each sharp discontinuity in the 
pulse shape corresponds to the arrival of
a photoelectron. (b) Example of experimental pulse recorded with the same 
$\tau_{int}$.}}
\label{MCexp}
\end{center}
\end{figure}

In Fig.~\ref{MCexp} a typical pulse generated by MC with 
$\tau_{int}=500$~ns is compared with an experimental pulse recorded 
with the same integration time.
The agreement between the two shapes is quite good.

In Fig.~\ref{figsigmaAsuA}b the dependence of 
$\sigma_A/\overline{A}$ and in Fig.~\ref{T} the mean value of the peaking 
time distribution ($\overline{T}$) on $\tau_{int}$, calculated with the MC,
are compared with the experimental points. In both cases the agreement is 
quite good. This confirms that in the present experimental conditions 
the experimental resolution is better than predicted by the naive model
based on a Poissonian statistcs. 
These checks give confidence in the MC simulation and allow to
use it to predict, in the most general experimental situation, which is
the energy resolution attainable with an integrator followed by a
peak-sensitive electronics.

\vskip 5 truemm

\subsection{Energy resolution in the general case}

\begin{figure}[b]
\begin{center}
\includegraphics[width =.85\textwidth] {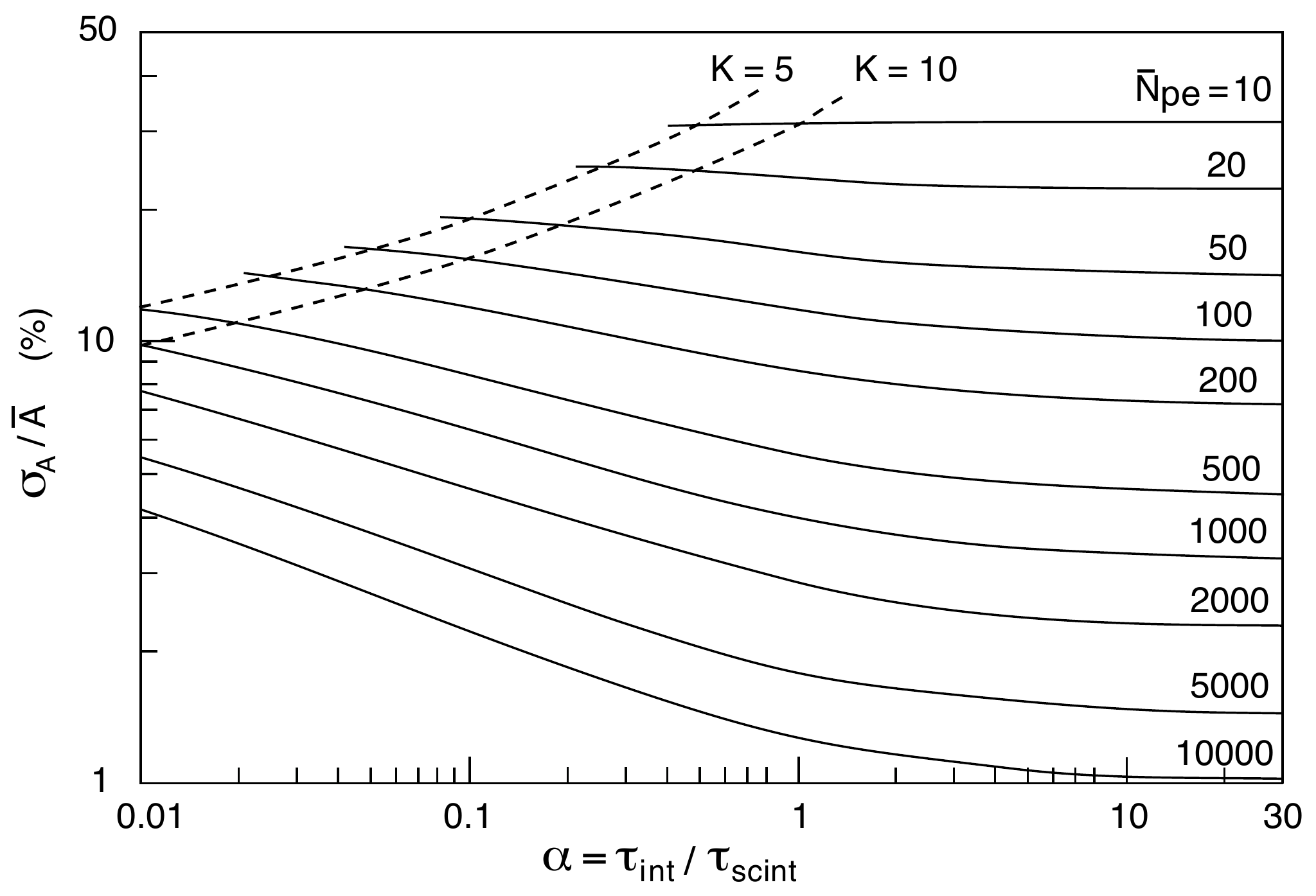}
\caption{\small{Look-up figure, calculated with the MC simulation,
which allows to  evaluate the maximum amplitude resolution for different
values of the integration time, and for different values of 
$\overline{N}_{pe}$. Only the region where $K \gtrsim 5$ 
(see Eq.\protect\ref{k}) is shown. 
}}
\label{lookuptable}
\end{center}
\vskip -2truemm 
\end{figure}

The resolutions $\sigma_A/\overline{A}$, calculated 
as a function of $\alpha\;=\;\tau_{int}/\tau_{scint}$
and for different values of
$\overline{N}_{pe}$, are reported in 
Fig.~\ref{lookuptable} which is therefore a general utility to evaluate
the resolution attainable with a scintillator having a decay time 
$\tau_{scint}$, read by a photodetector followed by an integrator and 
a peak-sensitive electronics. 

To present these results in 
a general form the fluctuations on the photodetector gain have not been 
included because they depend on the particular type used. For a 
PM  the effect of these fluctuations can be taken into account by 
multiplying the values of $\sigma_A/\overline{A}$ read on 
Fig.~\ref{lookuptable} by the corrective factor of
Eq.~\ref{eqsigmaEsuEOrig}.
From Fig.~\ref{lookuptable} it appears that to perform an integration with
$\tau_{int} < \tau_{scint}$, at least 10 photoelectrons are needed.

As already pointed out in the naive model, only the $n_{pe}$ 
photoelectrons emitted 
before the peaking time $T$ contribute, for each event, to the the maximum
amplitude $A$. Assuming a Poisson distribution for $n(T)$ the relative 
fluctuations on $A$ is given by:
 
\begin{equation} 
\frac{\sigma_A}{\overline{A}} = 
\frac{1}{\sqrt{\overline{n}_{pe}}} 
\label{naive}
\end{equation}
%

\begin{figure}[t]
\begin{center}
\includegraphics[width =.9\textwidth] {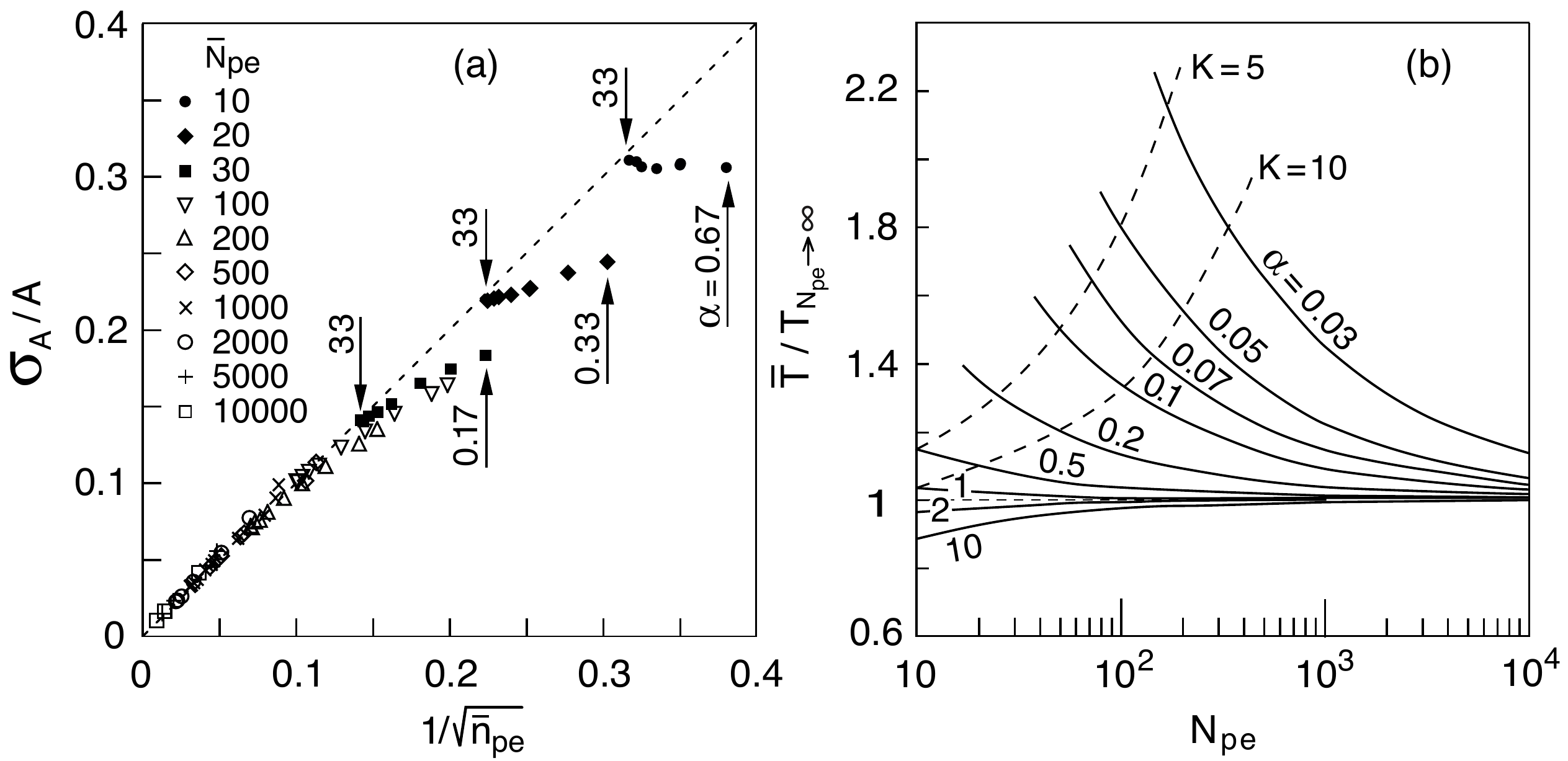}
\caption{\small{
(a) Relative maximium amplitude resolution as a 
function of the inverse square-root of the average number of 
photoelectrons which contribute to the maximum amplitude of the signal.
For the first and last points of the upper three series the value of 
$\alpha = \tau_{int}/\tau_{scint}$ is reported.
(b) Average peaking time divided by the peaking time at 
$\overline{N}_{pe} \to \infty$ (Eq.~\protect\ref{eqT}) as a function of 
the actual number of photoelectrons in each pulse, for different values of 
$\alpha$. Only the region where $K \gtrsim 5$ 
(see Eq.\protect\ref{k}) is shown.
}}
\label{figab}
\end{center}
\end{figure}

Contrarily to the experimental situation, in the MC simulation $n_{pe}$ is 
a known quantity for each event, so that the simple model represented 
by Eq.~\ref{naive} can be tested. 
In Fig.~\ref{figab}a $\sigma_A / \overline{A}$, calculated with the MC, is 
reported as a function of $1/\sqrt{\overline{n}_{pe}}$, 
for different values of $\overline{N}_{pe}$ and $\alpha $.
It appears that for $\overline{N}_{pe}\gtrsim 500$ Eq.~\ref{naive} 
is satisfied for any of the considered 
$\alpha $ values so that
the statistics of $n_{pe}$ is Poissonian and the naive model is valid. 
For smaller values of $\overline{N}_{pe}$ and in the range where
Eq.~\ref{k} is satisfied, the resolution is better than 
predicted by Eq.~\ref{naive} so that in that region the statistics is 
\hbox{sub-Poissonian}. 
The reason of this behaviour is clear from the curves reported
in Fig.~\ref{figab}b: for a fixed integration time 
a positive (negative) variation of $N_{pe}$ with respect of its
average value $\overline{N}_{pe}$ results in a negative (positive)
variation of the corresponding peaking time, which partially compensates
the variation on $N_{pe}$.
This anticorrelation between $N_{pe}$ and $T$ is responsible for the 
sub-Poissonian fluctuations at the lower values of $\overline{N}_{pe}$.

The results reported in Fig.~\ref{lookuptable} and Fig.~\ref{figab}
are valid for any integration time and for any decay time of the 
scintillating light, when the readout electronics measures the
maximum amplitude of the integrated pulse. 
\vspace*{4truemm} 

\section{Conclusions}
\label{conclusions}
The possibility of using a scintillating crystal with 
a slow decay time (like BGO)
for an electromagnetic calorimeter in a high-rate experiment was 
investigated. In these experimental conditions a fast measurement
of the energy deposited in the crystal is needed. This can be obtained,
at the cost of a lower energy resolution,
by integrating the output signal of the photodetector over a short time 
and by acquiring the maximum amplitude of 
the integrated signal. 

An experimental test and a Monte Carlo simulation show that the 
energy resolution comes from the statistics of the
number of photoelectrons emitted before the peaking time of the integrated 
pulse. While for a large number of 
photoelectrons the statistics follows 
a Poisson distribution, at a lower number of photoelectrons the statistics
becomes \hbox{sub-Poissonian} due to an anticorrelation between the 
fluctuations of the number of photoelectrons per pulse and the peaking 
time of that pulse.   
The results are reported in a general form which allows to evaluate the
contribution of the photoelectron statistics to the resolution
of a calorimeter equipped with a  
scintillating crystal read by a photomultiplier, followed by an 
integrator and a peak-sensitive electronics.
%

\end{document}